\documentclass{emulateapj}
\usepackage{gensymb}
\usepackage{upgreek}
\usepackage{color}
\def\spitz{{\it Spitzer}}

\def\wise{{\it WISE}}

\begin{document}

\shortauthors{Esplin \& Luhman}
\shorttitle{Survey for Stars and Brown Dwarfs in Ophiuchus}

\title{A Survey for New Stars and Brown Dwarfs in the Ophiuchus Star-forming
 Complex}

\author{
T. L. Esplin\altaffilmark{1,2}
and
K. L. Luhman\altaffilmark{3,4}
}

\altaffiltext{1}{Steward Observatory, University of Arizona, Tucson, AZ, 85719, 
USA; taranesplin@email.arizona.edu}
\altaffiltext{2}{Strittmatter Fellow}
\altaffiltext{3}{Department of Astronomy and Astrophysics, The Pennsylvania
State University, University Park, PA 16802; taran.esplin@psu.edu.}
\altaffiltext{4}{Center for Exoplanets and Habitable Worlds,
The Pennsylvania State University, University Park, PA 16802.}

\begin{abstract}
We have performed a survey for new members of the Ophiuchus cloud complex
using high-precision astrometry from the second data release of {\it Gaia},
proper motions measured with multi-epoch images from the {\it Spitzer Space
Telescope}, and color-magnitude diagrams constructed with photometry
from various sources.
Through spectroscopy of candidates selected with those data, we have
identified 155 new young stars. Based on available measurements of
kinematics, we classify 102, 47, and six of those stars as members of
Ophiuchus, Upper Sco, and other populations in Sco-Cen, respectively.
We have also assessed the membership of all other stars in the vicinity of
Ophiuchus that have spectroscopic evidence of youth from previous studies,
arriving at a catalog of 373 adopted members of the cloud complex.
For those adopted members, we have compiled mid-IR photometry
from {\it Spitzer} and the {\it Wide-field Infrared Survey Explorer}
and have used mid-IR colors to identify and classify circumstellar disks.
We find that 210 of the members show evidence of disks, including 48 disks
that are in advanced stages of evolution.
Finally, we have estimated the relative median ages of the populations
near the Ophiuchus clouds and the surrounding Upper Sco association
using absolute $K$-band magnitudes ($M_K$) based on {\it Gaia} parallaxes.
If we adopt an age 10~Myr for Upper Sco, then the relative values of $M_K$
imply median ages of $\sim2$~Myr for L1689 and embedded stars in L1688,
3--4~Myr for low-extinction stars near L1688, and $\sim6$~Myr
for the group containing $\rho$~Oph.

\end{abstract}

\keywords{accretion, accretion disks - brown dwarfs - protoplanetary disks -
stars: formation - stars: low-mass - stars: pre-main sequence}

\section{Introduction}

The Ophiuchus complex of dark clouds is one of the nearest and most studied
sites of active star formation \citep[$d\sim$140~pc,][]{ori17,ori18}.
It overlaps with the eastern edge of the Upper Scorpius subgroup 
\citep[10~Myr;][]{pec12,pec16} in the Scorpius-Centaurus OB association
and contains $\sim$300 known members \citep{wil08}. 
Separate populations among those members exhibit ages ranging from
$\lesssim1$~Myr for L1688 to several Myr for the group associated
with the star $\rho$~Oph \citep{gre95,luh99,pil16}.
The current census of Ophiuchus members has resulted from surveys spanning
more than 40 years, the first of which searched for stars with signatures of
youth in the form of H$\alpha$, X-ray, and mid-infrared (IR) emission
\citep[][references therein]{wil08}.
Later surveys incorporated photometry and proper motions from optical
and near-IR images \citep[][]{luh99,eri11}, extending to progressively 
fainter magnitudes and lower masses \citep{alv10,alv12,gee11,muz12}.

To work toward a more complete census of the stellar and substellar members of Ophiuchus,
we have performed a new survey of the region that is based primarily
on high-precision astrometry from the {\it Gaia} mission \citep{per01,deb12} 
and proper motions measured with multi-epoch imaging from the 
{\it Spitzer Space Telescope} \citep{wer04}.
In this paper, we describe the current census of Ophiuchus (Section~\ref{sec:cat}),
our identification of candidate members via photometry, proper motions, and parallaxes (Section~\ref{sec:ident}),
and our spectroscopic classification of candidate members (Section~\ref{sec:spec}).
In addition, we determine whether the known members host circumstellar disks 
and classify the evolutionary stages of any detected disks (Section~\ref{sec:disks}). 
Finally, we estimate the age differences among the 
populations projected against Ophiuchus (Section~\ref{sec:ages}). 

\section{Catalog of Known Members of Ophiuchus}
\label{sec:cat}

\begin{figure}[h]
	\centering
	\includegraphics[trim = 0mm 0mm 0mm 0mm, clip=true, scale=.6]{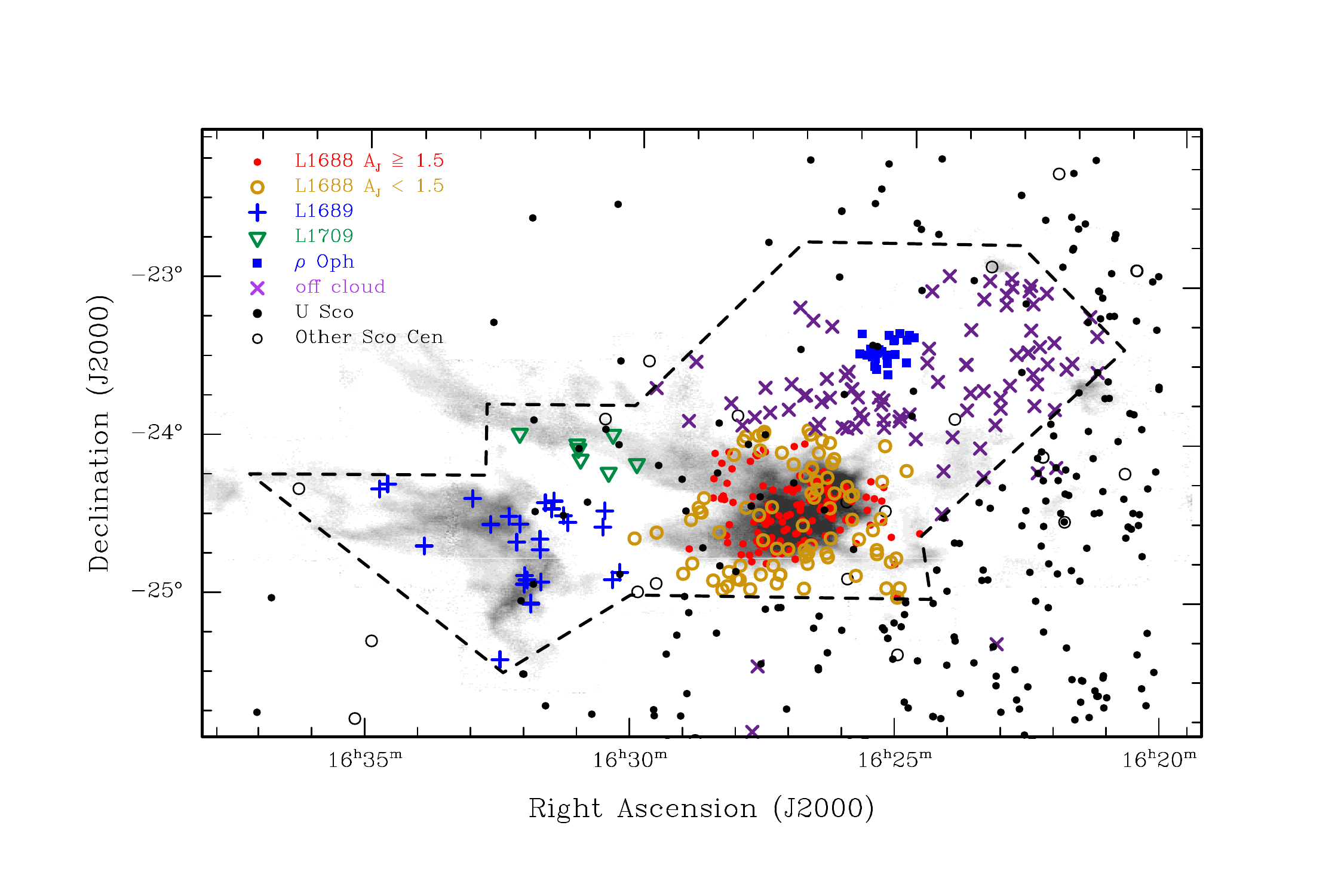}
\caption{
Spatial distribution of known young stars in the vicinity of the Ophiuchus complex.
Our adopted members of Ophiuchus (Table~\ref{tab:mem}) are assigned symbols based
on positions toward L1688 (red filled circles and yellow open circles),
L1689 (blue plus signs), L1709 (green open triangles), 
the $\rho$ Oph cluster (blue filled squares),
and areas without dark clouds (purple crosses).
Young stars that are not adopted Ophiuchus members consist of 
members of Upper Sco (black filled circles) and other populations in
Sco-Cen (black open circles). 
The dashed line indicates the boundary between Ophiuchus and Upper Sco 
from \cite{esp18}. The dark clouds of Ophiuchus are represented by a map of 
$^{13}$CO emission \citep{com}.}
\label{fig:regions}
\end{figure}

We have compiled all known objects with evidence of youth  
and measured spectral types within the boundary of Ophiuchus defined
by \citet{esp18}, many of which are taken from previous catalogs of
Ophiuchus members \citep[][]{mar98,wil08,mcc10,eri11,rig16}.
\citet{luh19} constructed a similar catalog for
an area outside of that boundary and within Upper Sco.
The Ophiuchus boundary from \citet{esp18} was defined to encompass a region
in which the average offset from the median $M_K$ of Upper Sco as a function
of spectral type was significantly brighter, which is an indication of
a younger average age.
That region includes the three main dark clouds of Ophiuchus and a
substantial area north of L1688, as shown in Figure~\ref{fig:regions}.

\begin{figure}[h]
	\centering
	\includegraphics[trim = 0mm 0mm 0mm 0mm, clip=true, scale=.6]{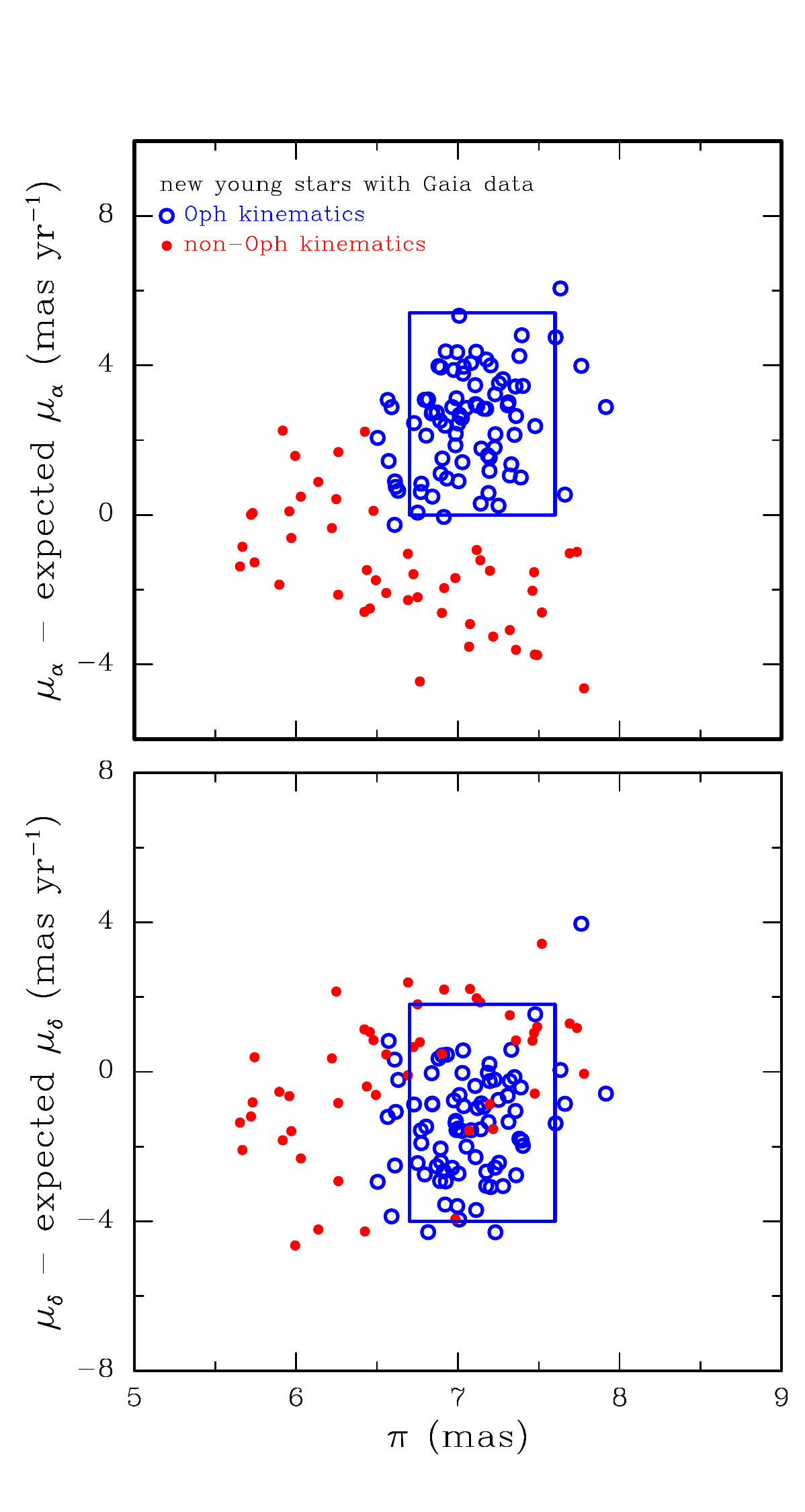}
\caption{
Proper motion offsets versus parallax for new young stars from
our spectroscopic sample that have parallaxes and proper motions from
{\it Gaia} DR2. We have indicated our adopted thresholds for membership
in Ophiuchus \citep[rectangles,][]{luh19}.
Stars with 1~$\sigma$ errors that overlap with those thresholds are
adopted as Ophiuchus members (blue open circles). The remaining stars have
kinematics that are consistent with membership in Upper Sco or other
populations in Sco-Cen (red closed circles).
The stars have been assigned symbols based on whether they satisfy those
criteria (at 1$\sigma$).
The proper motion offsets are relative to the values expected for the
positions and parallaxes of the stars assuming a space velocity of $U=-5$,
$V=-16$, and $W=-7$~km~s$^{-1}$.
Errors in parallax and proper motion offsets range from 0.03--0.7 mas and 0.06--1.63 mas/yr,
respectively. 
}
\label{fig:gaiapm}
\end{figure}

Our sample of young stars toward Ophiuchus may contain both objects
associated with the cloud complex and members of other populations
in Sco-Cen, particularly Upper Sco.
To help separate Ophiuchus members from other populations,
we apply the kinematic criteria from \cite{luh19} to data from
the second data release (DR2) of the {\it Gaia} mission \citep{gaia18}.
\citet{luh19} used data from that survey to characterize the parallaxes and
proper motion offsets ($\Delta \mu_{\alpha}$, $\Delta \mu_{\delta}$) for
members of Ophiuchus, Upper Sco, and the remainder of the Sco-Cen complex.
The proper motion offsets were defined as the difference between the observed 
proper motion and the motion expected at
the star's celestial coordinates and parallax if it had
the median space velocity of members of Upper Sco
($U,V,W = -5,-16,-7$ km s$^{-1}$). 
\citet{luh19} calculated density maps in $(\pi,\Delta\mu_{\alpha})$ and
$(\pi,\Delta\mu_{\delta})$ for young stars projected against Ophiuchus.
We have adopted thresholds for membership in Ophiuchus that approximate
the 10\% contours in those parameters. The resulting thresholds are shown in
Figure~\ref{fig:gaiapm}.
We adopt as members of Ophiuchus all known young stars that are
within the boundary of Ophiuchus from \citet{esp18}, have {\it Gaia}
parallaxes with errors of $\leq$10\% and have 1~$\sigma$
errors in parallax and proper motion offsets that overlap with those criteria.
We also assign membership to young stars outside of that spatial boundary
that satisfy the kinematic criteria for Ophiuchus but no other population
in Sco-Cen.
Stars that lack precise parallaxes from {\it Gaia} are included in our 
membership catalog if 
they are within the Ophiuchus boundary, show spectroscopic evidence of youth,
and do not have proper proper motion measurements that are inconsistent
with membership.
Due to the overlap in parallaxes and proper motions of members
of Ophiuchus and other populations in Sco-Cen \citep{luh19}, some of the
stars in our catalog for Ophiuchus could be members of the latter.

SR~21~A and B, $\rho$ Oph B, Gaia DR2 6049163679218606720 
do not satisfy our proper motion criteria,
but they are retained as members. The first two stars exhibit
evidence of youth and significant reddening, indicating that they are
associated with the Ophiuchus clouds. $\rho$~Oph~B and Gaia DR2 6049163679218606720
are likely
companions to $\rho$~Oph and GSS 37 given that they comprise pairs 
separated by only $3\arcsec$ and $1.1\arcsec$, respectively.

We have adopted 373 young stars as members of Ophiuchus, 
102 of which are newly classified in this work (Section~\ref{sec:spec}). 
The catalog is presented in Table~\ref{tab:mem}, which includes
celestial coordinates, available measurements of spectral types, our
adopted types, {\it Gaia} DR2 identification numbers, and the 
coordinate-based designations from the Point Source Catalog of 
the Two Micron All Sky Survey \citep[2MASS;][]{skr06},
the tenth data release of the United Kingdom Infrared Telescope Infrared Deep
Sky Survey \citep[UKIDSS,][]{law07}, and the AllWISE Source Catalog
from the {\it Wide-field Infrared Survey Explorer} \citep[{\it WISE};][]{wri10}.
Table~\ref{tab:mem} also contains astrometry and optical photometry from
{\it Gaia} DR2, near- and mid-IR photometry from
various sources, disk classifications, and extinction estimates.
One of the {\it Gaia} parameters is the renormalized unit weight error 
\citep[RUWE,][]{lin18}, which is an indicator of the quality of the
astrometric fit.

In Table~\ref{tab:mem}, we have assigned each adopted member of Ophiuchus 
to one of five groups based on their locations:
objects projected against the dark clouds L1688, L1689, and L1709,
objects in the vicinity of the star $\rho$~Oph,
and the remaining off-cloud stars. 
These assignments are also indicated in the map in Figure~\ref{fig:regions},
where we have further divided the L1688 population based on extinctions
above and below $A_J=1.5$ ($A_K=0.57$) (Section~\ref{sec:ext}).
Using the available parallaxes with $\sigma_\pi/\pi< 0.1$ from
{\it Gaia} DR2 for our adopted members, we measure median distances
of 138, 145, 137, and 140~pc for the groups toward L1688, L1689, L1709,
and $\rho$~Oph, respectively.

In Table~\ref{tab:other}, we present 59 young stars 
that have spectral classifications from previous studies
or this work, are within the boundary of Ophiuchus complex, and
exhibit kinematics inconsistent with membership in Ophiuchus.
The catalog includes the available spectral classifications, photometry
and astrometry from {\it Gaia} DR2, near-IR photometry from 2MASS,
and flags indicating whether the {\it Gaia} astrometry is consistent with
membership in Upper Sco or the remainder of Sco-Cen \citep{luh19}.

\section{Identification of Candidate Members}
\label{sec:ident}

We have searched for new candidate members of Ophiuchus using parallaxes,
proper motions, and optical and near-IR color-magnitude diagrams from
several sources. In this section, we describe the procedures for the
identification of those candidates.

\subsection{Proper Motions and Parallaxes from {\it Gaia} DR2}

The high-precision astrometry from {\it Gaia} DR2 has greatly improved
the ability to identify members of nearby associations
\citep[e.g.,][]{gag18d,luh18tau,dam19,her19}.
Among the 271 stars that we have adopted as previously known members
of Ophiuchus, parallaxes with $\leq10$\% errors are available from {\it Gaia}
DR2 for 193 objects, which have spectral types as late as M8.0. 
We have searched for new members of Ophiuchus within our
adopted boundary for the cloud complex by applying the kinematic criteria
described in the previous section to astrometry from {\it Gaia} DR2.
As done for a survey of Upper Sco by \citet{luh19}, we applied the
criteria to {\it Gaia} sources with $\pi/\sigma\geq10$ and have ignored the
{\it Gaia} astrometry if it is inconsistent with membership and may be
unreliable based on a large value of RUWE ($>1.6$).
These criteria produce 168 candidate Ophiuchus members that  
lack definitive assessments of youth from previous spectra.

\citet{can19} recently identified 166 candidate members of Ophiuchus 
using astrometry from {\it Gaia} DR2.
Among their candidates, 75 are also within our sample, four are slightly
inconsistent with our proper motion criteria,
and the remainder are either located outside of our 
adopted boundary for Ophiuchus or have previously reported spectral
classifications.

\subsection{Proper Motions from the Spitzer Space Telescope}

Because {\it Gaia} observes at optical wavelengths, it is not sensitive
to Ophiuchus members that have high extinctions or very low masses.
Detection of such objects requires IR data.
Proper motions can be measured at IR wavelengths using the 
multiple epochs of imaging that are available for Ophiuchus from the
Infrared Array Camera \citep[IRAC;][]{faz04} on the {\it Spitzer
Space Telescope}.
From August 2003 to May 2009, IRAC operated with
four broad-band filters centered at 3.6, 4.5, 5.8, and 8.0~$\mu$m (denoted
as [3.6], [4.5], [5.8], and [8.0]) and four 256$\times$256 arrays.
The plate scale for each array was $1\farcs2$ pixel$^{-1}$, corresponding
to a field of view of $5\farcm2\times5\farcm2$.
Point sources exhibited a FWHM of $1\farcs6$--$1\farcs9$ for [3.6]--[8.0]. 
After the depletion of the cryogens, IRAC continued to function with the
[3.6] and [4.5] bands until January 2020.

\begin{figure}[h]
	\centering
	\includegraphics[trim = 0mm 0mm 0mm 0mm, clip=true, scale=.6]{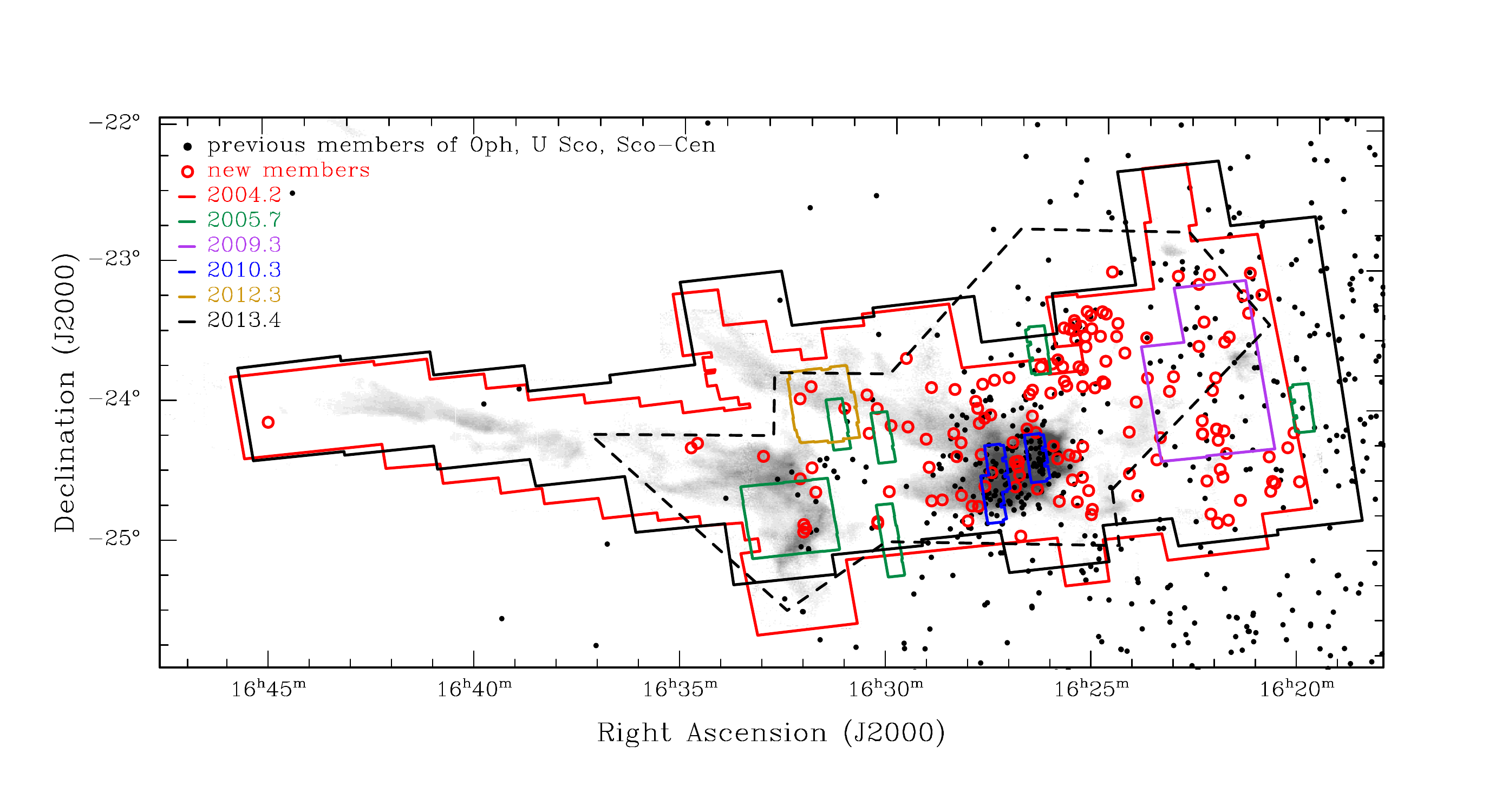}
\caption{
Map of the largest fields in Ophiuchus that were imaged by IRAC 
(Table~\ref{tab:epochs}), which span a wide range of epochs.
We have included the previously known members of Ophiuchus, Upper Sco, and Sco Cen
near the Ophiuchus dark clouds (black filled circles) 
and members discovered in this work (red open circles). 
}
\label{fig:iraccoverage}
\end{figure}

We have retrieved all [3.6] and [4.5] images in the vicinity of Ophiuchus
from the \spitz\ archive. The Astronomical Observing Requests (AORs), program 
identifications (PIDs), and principles investigators (PIs)
for those observations are listed in Table~\ref{tab:epochs}.
The images within a given {\it Spitzer} observing window were combined, and
they correspond to a single epoch in our analysis. The observations in each
epoch spanned $\lesssim1$~month. These IRAC data have been used for studying 
outflows from young stars \citep{sea08},
estimating star formation efficiencies and disk frequencies \citep{eva09},
searching for brown dwarfs with circumstellar disks \citep{har10},
confirming the nature of brown dwarfs selected by \wise\ \citep{gri12},
measuring the variability of young stars \citep{gun14},
and constraining the properties of dust in molecular clouds \citep{lef14}.
In Figure~\ref{fig:iraccoverage}, we show the fields that were observed
in the six programs that covered the largest areas.
Most of Ophiuchus was observed by programs 177 and 90071, which were separated 
by nine years. Those programs obtained two and nine 10.4~s exposures, 
respectively, at each position and filter.

We measured astrometry from the [3.6] and [4.5] images using
the methods from \citet{esp17}.
In summary, we performed the following steps for the 
final epoch of observations:
1) measured pixel coordinates, fluxes ($F_\nu$), and signal-to-noise rations (SNRs) 
for all sources in each image using the point-response-function
fitting routine in the Astronomical Point source Extractor \citep[APEX, ][]{mak05};
2) applied the distortion correction from \cite{esp16} to the pixel coordinates,
3) estimated the central coordinates and orientations of each image using astrometry from {\it Gaia} DR2,
4) iteratively refined those coordinates and orientations
and 5) calculated celestial coordinates for each detection and averaged the detections from both bands
of a single source. 
To avoid unreliable astrometry due to low SNR and saturation,
we ignored sources with less than three total detections in a given epoch
across the two bands and individual detections with 
$F_\nu /({\rm exposure\ time})>0.73$ and $>$0.82~Jy/s in [3.6] and [4.5],
respectively. For the other epochs,
the central coordinates and orientations of images were aligned to the 
astrometry of the final epoch. 

\begin{figure}[h]
	\centering
	\includegraphics[trim = 0mm 0mm 0mm 0mm, clip=true, scale=1]{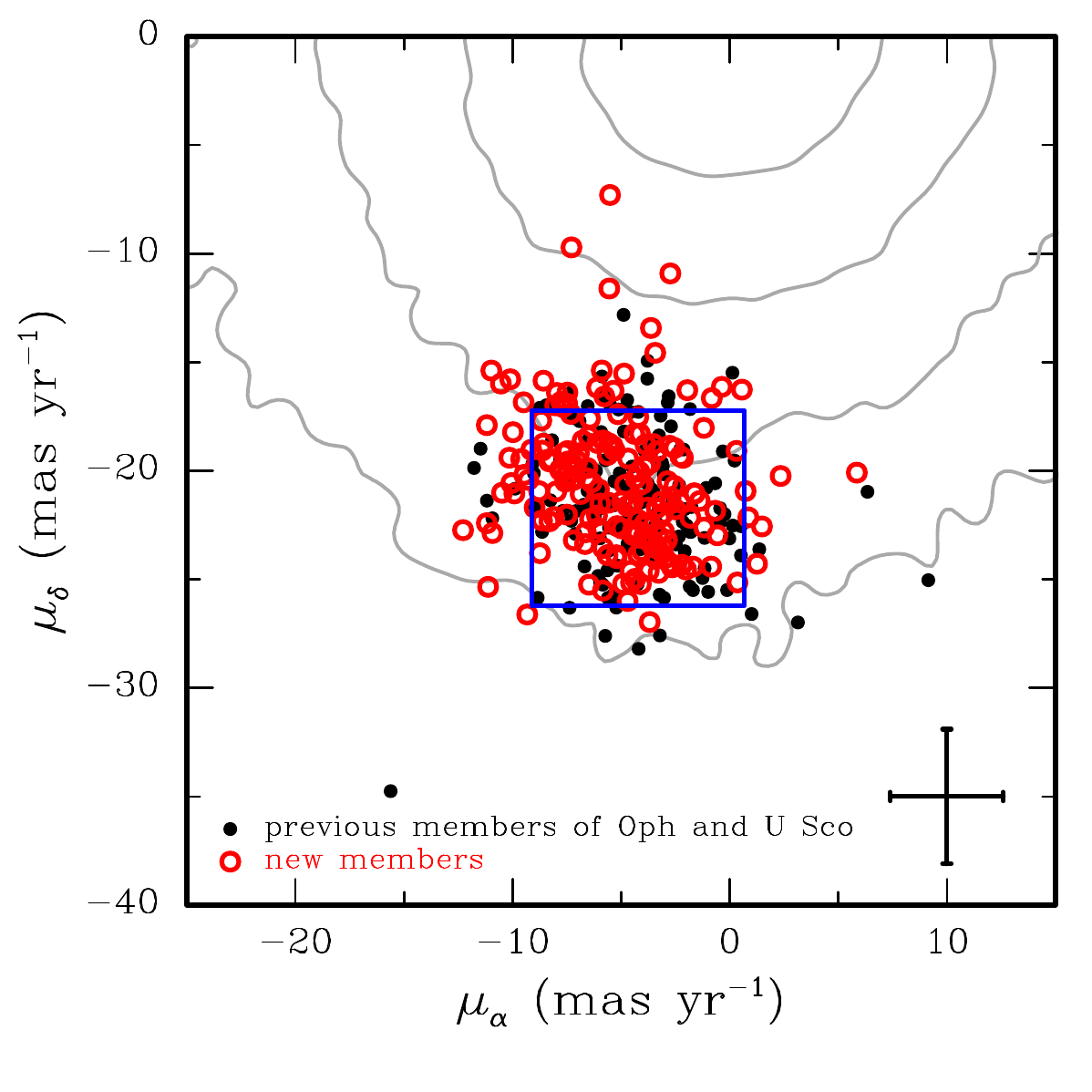}
\caption{
Relative proper motions measured from multi-epoch IRAC imaging
for previously known members of Ophiuchus and Upper Sco (black filled circles)
and new members discovered in this work (red open circles).
Measurements for other sources in the IRAC images are represented
by contours at log(numbers/(mas yr$^{-1}$)$^2$) = 1.0, 1.5, 2.0, and 2.5.
The typical errors are indicated (error bars).
}
\label{fig:iracpm}
\end{figure}

Proper motions were calculated for each source using a linear fit of 
the right ascension and declination as a function of time. 
In our analysis of proper motions,
we ignored sources with errors in $\mu_{\alpha}$ or $\mu_{\delta}$ $>$10~mas~yr$^{-1}$
and SNRs $\leq$4.5 and 5.5 in [3.6] and [4.5], respectively. 
These thresholds were determined in the manner described by \cite{esp17}.
In Figure~\ref{fig:iracpm}, we show the relative proper motions measured
with IRAC for previously known members of Ophiuchus and new members
from this work.
The IRAC proper motions are included in Table~\ref{tab:mem}.

To identify IRAC sources that have proper motions consistent with membership
in Ophiuchus, we have applied thresholds that span the same ranges of values as
the criteria used for {\it Gaia} DR2 and that are centered on the median
IRAC motion of known members, as shown in Figure~\ref{fig:iracpm}.
These criteria produce a sample of roughly 2000 candidates that are not
previously known members. Most of the candidates are rejected via
the photometric criteria described in the next section.
Since the IRAC images extend beyond our adopted boundary for Ophiuchus
(Figure~\ref{fig:iraccoverage}), some of the proper motion candidates do so
as well.
Eight previously known members of Ophiuchus have IRAC proper motions
that do not satisfy our criteria for membership. Seven of those stars exhibit
enough extinction ($A_V>4$) and evidence of youth to indicate that
that they are associated with the Ophiuchus clouds, so we retain them
as members.
The one remaining object, 2MASS J16265128-2432419, has lower extinction,
but its {\it Gaia} data are consistent with membership, so we adopt it
as a member as well.

\subsection{Color-magnitude Diagrams} 
\label{sec:photo}

To further refine the $>$2000 Ophiuchus candidates selected with astrometry
from {\it Gaia} and {\it Spitzer},
we have checked whether the candidates have positions in 
optical and near-IR color-magnitude diagrams (CMDs) that are 
similar to those of the previously known members of Ophiuchus.

We have searched for counterparts 
of the previously known members of Ophiuchus and the candidate members
selected with {\it Gaia} and IRAC in publicly available photometric catalogs
at optical and IR wavelengths. 
The data included in our analysis consist of: $G$, $G_{BP}$, $G_{RP}$ 
(3300-10500, 3300--6800, 6300--10500 \AA) from {\it Gaia} DR2;
$rizy_{P1}$ from the first data release of Pan-STARRS1 \citep[PS1,][]{kai02,kai10,fle16};
$ZYJHK$ from data release 10 of UKIDSS;
$JHK_s$ from the 2MASS Point Source Catalog;
and $JK_s$ from data release 6 of the VISTA Hemisphere Survey (VHS).
We adopted the point-spread function (PSF) 
magnitudes from the stacked images in PS1
and the 1$\arcsec$ radius aperture magnitudes from UKIDSS and VHS.
Potentially saturated measurements from these surveys were avoided in the
manner described by \citet{luh18}.

In addition to public surveys, we have made use of photometry that
we measured from the 3.6 and 4.5~\micron\ IRAC images and $JHK_s$ images
that we obtained with the Infrared Side Port Imager (ISPI) at the Cerro Tololo 
Inter-American Observatory (CTIO). 
We converted the 2013.4 epoch [3.6] and [4.5] $F_\nu$ measurements from 
APEX into magnitudes using zero points selected to align with IRAC 
photometry from \cite{luhM12}. 
The ISPI images were obtained on the nights of April 28--30 in 2004.
They cover a $20\arcmin\times20\arcmin$ field centered at 
$(\alpha,\delta) = (246.542\arcdeg,-24.377\arcdeg)$ and have 
completeness limits near $J=18.5$, $H=17.7$, and $K_s=16.7$.

We combined and merged the above catalogs of photometry in the manner
done for Upper Sco by \citet{luh18}.
UKIDSS and VHS exhibit color-dependent offsets relative to 2MASS, 
so we have added 0.07, $-$0.04, 0.03 mag to $J$, $H$, and $K$ from UKIDSS
and 0.07, $-$0.03 to $J$ and $K$ from VHS,
which aligns those data to the 2MASS photometry for known M-type members
of Upper Sco \citep{luh19}. 
In our analysis, photometric measurements with errors greater than 0.1~mag
are ignored.
We have estimated the extinction for each source by dereddening 
it to the typical locus of young stars at the distance of Ophiuchus
in $H$ versus $J-H$ (or $K_s$ versus $J-K_s$ when $J-H$ was not available),
as done in our previous work \citep[][references therein]{espl17}.
These extinctions are used to deredden the photometry of the
stars in our other CMDs, which helps to distinguish Ophiuchus members
from field stars. In Section~\ref{sec:ext}, we will compute separate
extinctions for the adopted members in a way that utilizes their measured
spectral types.

\begin{figure}[h]
	\centering
	\includegraphics[trim = 0mm 0mm 0mm 0mm, clip=true, scale=.6]{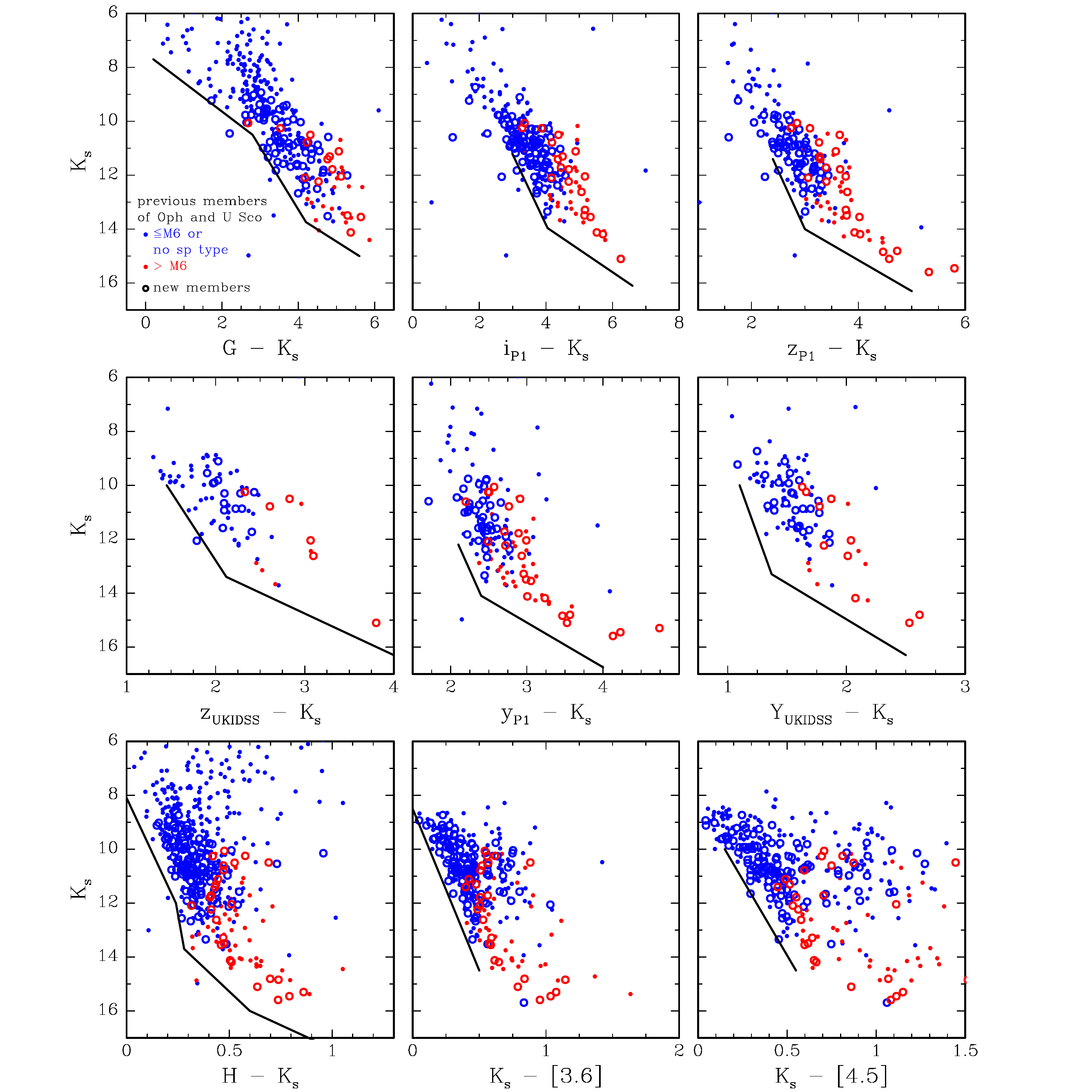}
\caption{
Extinction-corrected CMDs for previously known members of Ophiuchus and Upper Sco
within Figure~\ref{fig:iraccoverage} (filled circles) and new members from
this work (open circles).  Candidate members were identified based
on their positions above the solid boundaries.
}
\label{fig:criteria}
\end{figure}

In Figure~\ref{fig:criteria}, we present extinction-corrected CMDs for
previously known members of Ophiuchus and Upper Sco within 
Figure~\ref{fig:iraccoverage} and new members from this work.
As done in our previous surveys \citep{luh18}, 
each diagram has $K_s$ on the vertical axis and a color between
another band and $K_s$ on the horizontal axis. 
In each CMD, we have defined a boundary that follows the lower envelope of
known members.
Among the candidates identified with {\it Gaia} and IRAC astrometry,
405 sources appear above a boundary in at least one CMD and do not fall below
the boundaries of any CMDs, and thus are considered viable candidates.

\section{Spectroscopy of Candidate Members}
\label{sec:spec}

\subsection{Observations}

We have obtained spectra of 148 candidate members of Ophiuchus
from the previous section to confirm their membership via signatures of youth
and to measure their spectral types. In addition, we have performed spectroscopy
on 49 objects that were candidates early in our survey but that do not
satisfy our latest criteria and we have observed 33 previously known members
to improve their classifications, resulting in a total of 230 sources
in our spectroscopic sample.
There are 259 remaining candidates that lack either definitive 
assessments of youth from spectra or measured spectral types.
Those objects are are presented in Table~\ref{tab:cand}. 
We note that more than a third of the these candidates have been previously 
identified in the literature as potential or likely members of Ophiuchus.
For instance, 2MASS 16244941-2459388 exhibits evidence of youth from
previous spectroscopy \citep{rig16}. Since it lacks a measured spectral
type, we include it among the candidate members rather than the adopted
members.

Optical spectroscopy was performed with the Cerro Tololo Ohio State 
Multi-Object Spectrograph (COSMOS) at the 4-m Blanco telescope at CTIO and the 
Inamori Magellan Areal Camera and Spectrograph (IMACS) on the Magellan I
telescope at Las Campanas Observatory \citep{dre11}.
The former is based on an instrument described by \citet{mar11}.
Near-IR spectra were obtained with the
Cornell Massachusetts Slit Spectrograph \citep[CorMASS;][]{wil01}
on the Magellan II telescope,
the Astronomy Research using the Cornell Infra Red Imaging Spectrograph 
(ARCoIRIS) at the 4-m Blanco telescope at CTIO,
FLAMINGOS-2 on the Gemini South Telescope \citep{eik04},
and SpeX \citep{ray03} at the NASA Infrared Telescope Facility (IRTF).
The instrument configurations are summarized in Table~\ref{tab:log}.
The dates of the observations are listed in Table~\ref{tab:spec}.

\begin{figure}[h]
	\centering
	\includegraphics[trim = 0mm 0mm 0mm 0mm, clip=true, scale=.8]{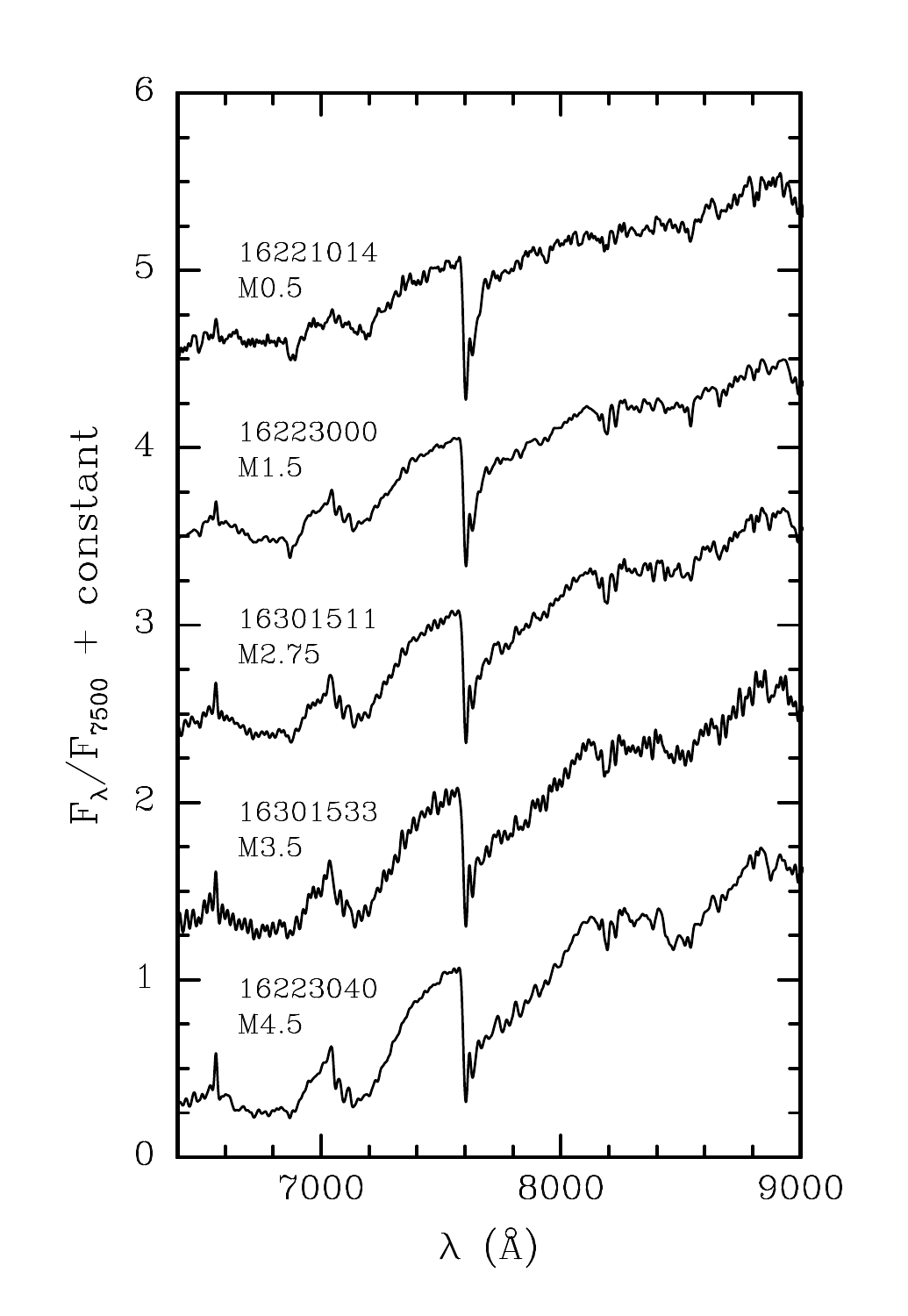}
\caption{
Examples of optical spectra of new members of Ophiuchus and Upper Sco
(Table~\ref{tab:spec}). These data are displayed at a resolution of 13 \AA.
The data used to create this figure are available.
}
\label{fig:optfig}
\end{figure}

\begin{figure}[h]
	\centering
	\includegraphics[trim = 0mm 0mm 0mm 0mm, clip=true, scale=.8]{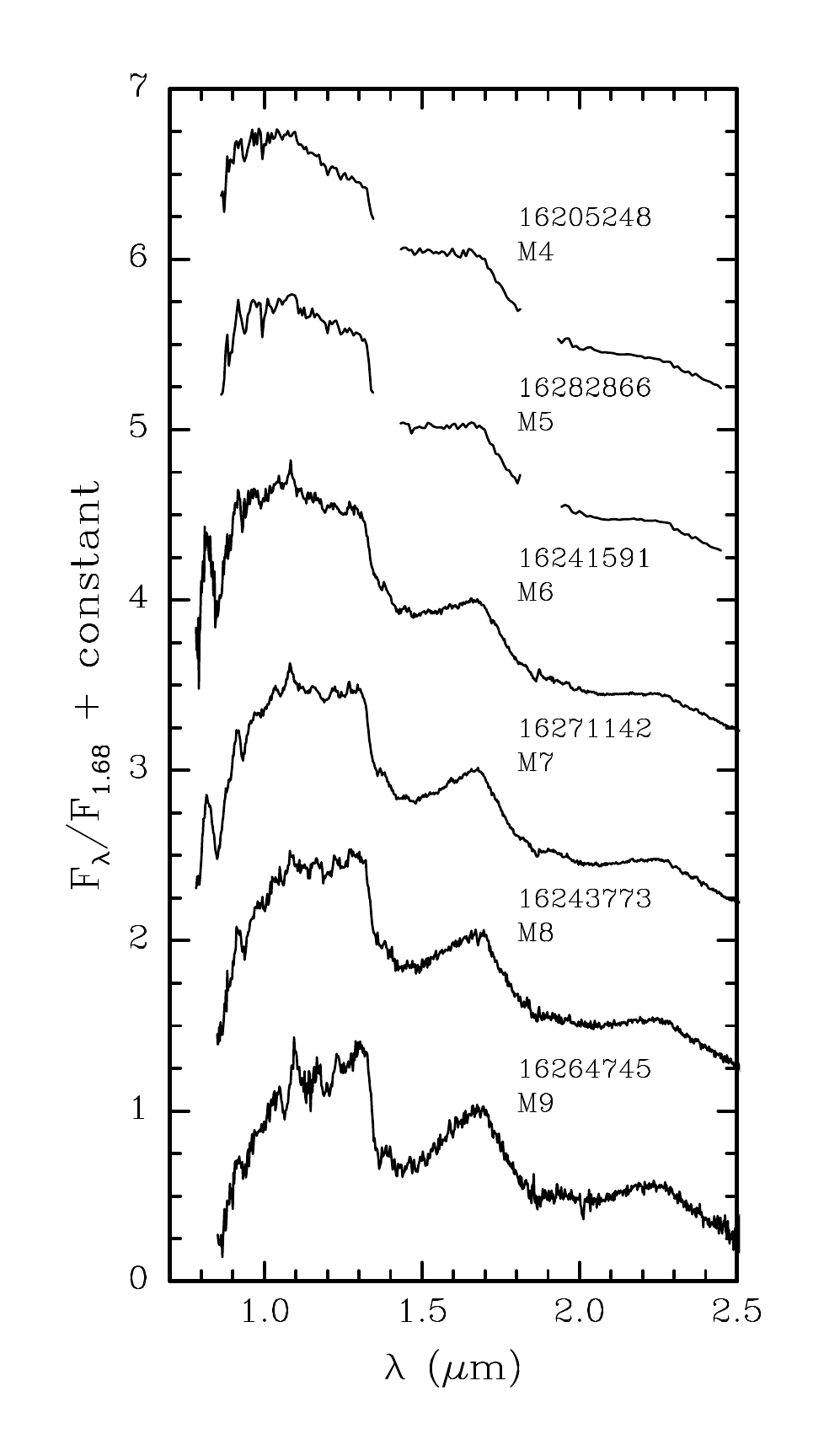}
\caption{
Examples of IR spectra of new members of Ophiuchus and Upper Sco 
(Table~\ref{tab:spec}). The spectra have been dereddened to match the slopes
of the young standards from \citet{luh17}.
The data used to create this figure are available.
}
\label{fig:irfig}
\end{figure}

We reduced the optical spectra using routines in IRAF.
The reduction steps included flat-field correction, spectral extraction,
and wavelength calibration.
The IR spectra from FLAMINGOS-2 and GNIRS were reduced in a similar 
manner with the addition of a step to correct for telluric absorption using 
a spectrum of an A-type star that was observed at a similar airmass.
Because the FLAMINGOS-2 spectra had low S/N, we binned the
spectra by a factor of 15.
The SpeX data were processed using the Spextool package \citep{cus04} and
corrected for telluric absorption in the manner from \citep{vac03}.
The ARCoIRIS spectra were reduced with a modified version of Spextool. 
Examples of the optical and IR spectra are presented in
Figures~\ref{fig:optfig} and \ref{fig:irfig}, respectively.
The reduced spectra are provided in electronic files associated with
those figures.

\subsection{Spectral Classification}\label{sec:class}

We have used our spectroscopic data to measure spectral types and
check for signatures of youth that would support membership in Ophiuchus.
The objects in our spectroscopic sample range from $K_s\sim9$--17, which
corresponds to spectral types of $\sim$M0 through early L for members
of Ophiuchus. At those types, we can distinguish young objects from field stars
via Li I absorption at 6707~\AA\ and gravity-sensitive features such as
Na~I and the steam absorption bands \citep{mar96,luh97,luc01}.
Spectral types were estimated for field stars using 
dwarf and giant spectral standards \citep{hen94,kir91,kir97,cus05,ray09}.
Optical spectral types of young objects were measured with averages
of spectra of dwarf and giants standards \citep{luh97,luh98,luhman99}.
IR spectra of young stars were classified using standard spectra 
constructed from optically-classified young objects \citep{luh17}. 

Our spectral classifications are presented in Table~\ref{tab:spec}.
We find that 159 of the objects in our spectroscopic sample 
that were not previously known members exhibit evidence
of youth. We have classified the membership for those young stars in the 
following manner. Stars are adopted as members of Ophiuchus if they have
1) locations within the Ophiuchus boundary and {\it Gaia} parallaxes and
proper motions consistent Ophiuchus \citep{luh19}, 2) locations
within Ophiuchus and no {\it Gaia} parallaxes or proper motions,
or 3) locations outside of Ophiuchus and {\it Gaia} parallaxes and 
proper motions that are consistent with Ophiuchus and no other populations
in Sco-Cen. If none of those sets of criteria are satisfied, then a young
object is assigned to Upper Sco if 1) its {\it Gaia} parallaxes and proper
motions are consistent with that population or 2) {\it Gaia} astrometry
is not available and it is located outside of Ophiuchus.
Stars that are not classified as members of Ophiuchus or Upper Sco
and that have astrometry consistent with the remainder of Sco-Cen \citep{luh19}
are assigned to that population.
In that way, we classify 102, 47, and six of the young sources 
without previously reported spectral types as new members
of Ophiuchus, Upper Sco, and the remainder of Sco-Cen, respectively.
Five young stars have kinematics that are inconsistent with
membership in any of the Sco-Cen populations.
In Table~\ref{tab:spec}, we indicate for each object whether it has
evidence of youth and, if so, its assigned population.
The spectral types of the new Ophiuchus members range from M1 to M9.5,
including 23 with spectral types later than M6 ($\lesssim0.1$~$M_\odot$).
Those objects also contain 6 of the 10 faintest known members of Ophiuchus
in terms of extinction-corrected $K_s$.

\section{Circumstellar Disks in Ophiuchus}
\label{sec:disks}

We have compiled mid-IR photometry for our catalog of 373 adopted members of
Ophiuchus from {\it WISE} and the {\it Spitzer Space Telescope}
and have used those data to identify the presence of disks and to
classify the detected disks.

\subsection{Mid-IR photometry}

\wise\ provides photometry in four bands 
centered at 3.4, 4.6, 12, and 22 \micron\ ($W1$--$W4$).
For the members of Ophiuchus, we have adopted the measurements from
the AllWISE Source Catalog when available.
Among members of Ophiuchus that lack data in that catalog,
five and two objects have counterparts in the \wise\ All-Sky Source Catalog
and the AllWISE Reject Table, respectively.
In total, \wise\ data are available for 349 members of Ophiuchus, 
which are included in Table~\ref{tab:mem}.
We visually examined the \wise\ images for each member to
check for spurious detections, blending with nearby sources, and
contamination from extended emission. We have flagged those data
accordingly in Table~\ref{tab:mem}.
We found 99, 94, 67, and 41\% of the members detected by \wise\
to have reliable photometry at $W1$, $W2$, $W3$, and $W4$, respectively. 

We also have made use of photometry from \spitz\ in the four bands of
IRAC and the 24 \micron\ band ([24]) of the Multiband Imaging Photometer
for {\it Spitzer} \citep[MIPS;][]{rie04}.
The latter produced images with a field of view of $5\farcm4\times5\farcm4$.
Point sources in those exhibits exhibited a FWHM of $5\farcs9$.
When available, we have adopted flux measurements
from the \spitz\ Enhanced Imaging Products (SEIP) Source List.
Those data were converted to magnitudes using zero points from \citet{esp18},
which were selected to match photometry from \cite{luhM12} for Upper Sco.
We inspected the SEIP mosaics to check for Ophiuchus members that were 
detected even though they lacked counterparts in the SEIP Source List
and to check for blends with neighboring stars.
For some of the latter cases, we performed PSF subtraction on the
contaminating objects and manually measured photometry for the targets
of interest. Aperture photometry for all detected objects without
SEIP measurements was measured using routines in the IDL's Astronomy Users
Library. We present \spitz\ photometry for 360 members of Ophiuchus in
Table~\ref{tab:mem}.
We have flagged the measurements that may be uncertain due to 
blending with neighboring stars or bright extended emission.

\subsection{Extinction Estimates}
\label{sec:ext}

We wish to determine whether each member of Ophiuchus exhibits mid-IR
excess emission from a circumstellar disk. Such excess emission can
be detected by comparing colors between near- and mid-IR bands to
those expected from stellar photospheres. However, because many members of
Ophiuchus are embedded in dark clouds, their colors are subject to
significant reddening from dust along the line of sight.
As a result, we must apply a correction for that reddening in order
to assess the presence of disk emission.
We have estimated extinction in the $K_s$ band ($A_K$) for
members of Ophiuchus based on the slope of near-IR spectra relative to the young
spectroscopic standards from \cite{luh17} when such data are available.
Otherwise, we have derived extinctions from color excesses in $J-H$
(or $H-K$ when $J$ is unavailable) relative to the intrinsic values
for young stars \citep{luh19} using the extinction law of \cite{sch16}.     
The resulting estimates of $A_K$ are included in Table~\ref{tab:mem}.

\subsection{Measurements of Excess Emission}

\begin{figure}[h]
	\centering
	\includegraphics[trim = 0mm 0mm 0mm 0mm, clip=true, scale=1]{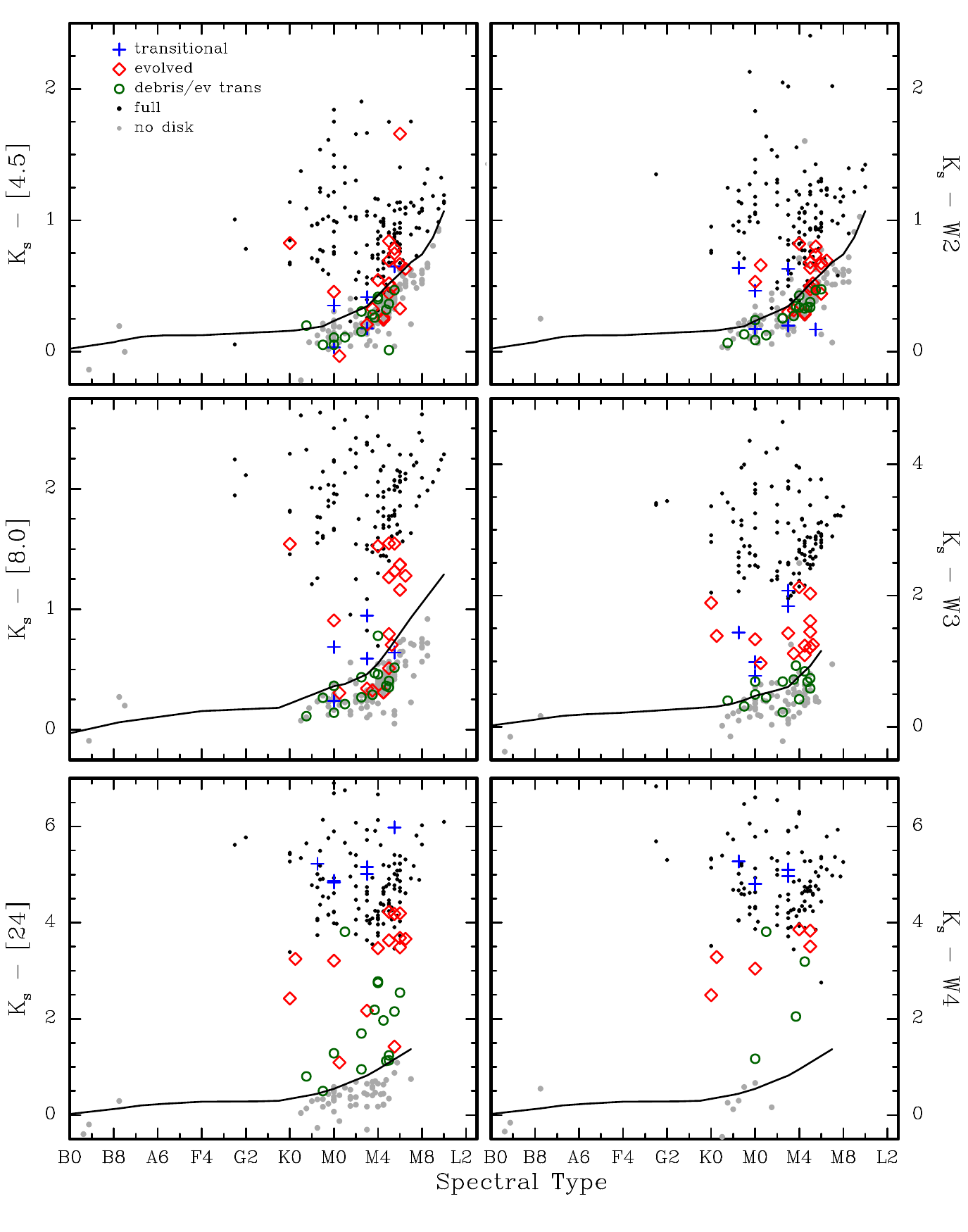}
\caption{
Extinction-corrected IR colors as a function of spectral type for known
members of Ophiuchus. 
In each color, we have marked a boundary that has been used to identify
the presence of color excess from circumstellar disks (solid lines), 
which were selected to follow the observed photospheric sequences 
\citep{esp18}.}
\label{fig:excess}
\end{figure}

We have used extinction-corrected mid-IR colors to identify excess emission
from disks among the adopted members of Ophiuchus.
In Figure~\ref{fig:excess}, we have plotted the colors between $K_s$ and
six bands from \wise\ and {\it Spitzer} as a function of spectral type.
The colors have been corrected for extinction using the values of $A_K$
from Table~\ref{tab:mem} and the extinction relations from \cite{ind05} and
\cite{sch16}.  
In each color, stars that lack disks form a blue sequence, corresponding
to the intrinsic photospheric colors, although the sequence is not 
well-populated in $K_s-W4$.
We have marked the thresholds for identifying excess emission 
from \cite{esp18}, which have been slightly shifted to reflect new
estimates of the intrinsic photospheric colors of young stars \citep{luh19}.
We provide flags in Table~\ref{tab:mem} indicating whether an object has excess 
emission in each of the six mid-IR bands in Figure~\ref{fig:excess}.
Sources with no apparent excess emission at the longest available wavelength 
were labeled as likely having no excess at all wavelengths. 
Excesses that are only slightly above our thresholds in $W3$, $W4$, [8.0],
and [24] are considered to be tentative.
Although the B9 star V2394 Oph shows excess emission in the mid-IR bands
relative to $K_s$, excesses are not present relative to $W1$ or [3.6], so
we classify it as having no excesses.

\subsection{Disk Classifications}

\begin{figure}[h]
	\centering
	\includegraphics[trim = 0mm 0mm 0mm 0mm, clip=true, scale=1]{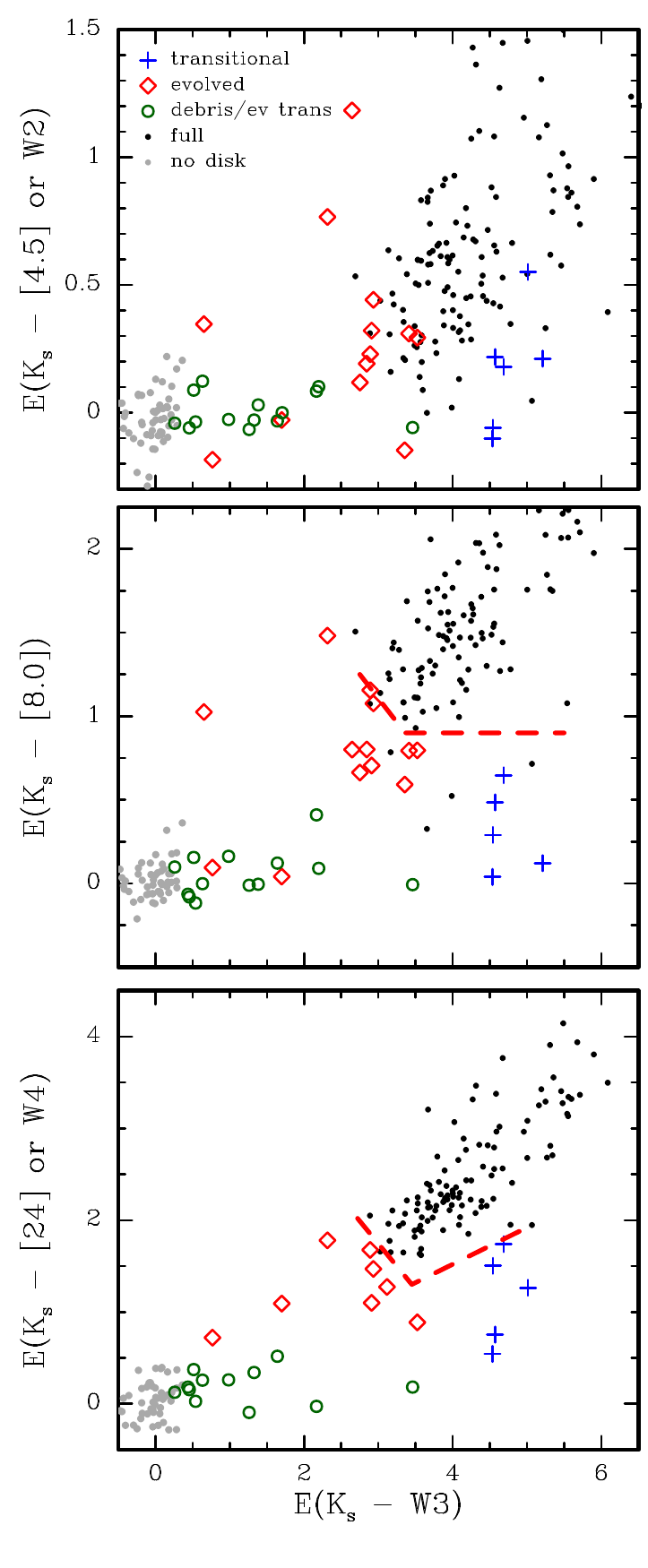}
\caption{
Extinction-corrected IR color excesses for known members of Ophiuchus. 
Data at [4.5] and [24] are shown when available. 
Otherwise, data from the similar bands of $W2$ and $W4$ are used.
In the bottom two diagrams, we indicate the boundaries that are used to 
distinguish full disks from disks in more advanced stages of evolution and 
are defined in \cite{esp14}.}
\label{fig:diskclass}
\end{figure}

For the 210 sources with excess emission in at least one band, 
we have classified the evolutionary stages of the circumstellar disks
using the methods that we have employed in Upper Sco 
\citep{luhM12,esp18}. 
We have considered the following disk classes
\citep[][]{ken05,rie05,her07,luh10,esp12}:
{\it full} (optically thick with no large holes),
{\it transitional} (optically thick with a large hole),
{\it evolved} (optically thin with no large hole),
{\it evolved transitional} (optically thin with a large hole),
and {\it debris disk} (second generational dust from planetesimal collisions). 
We have assigned these classes based on extinction-corrected color
excesses \citep{esp14,esp18}, as shown in Figure~\ref{fig:diskclass}.
The resulting classifications are listed in Table~\ref{tab:mem}. 
We classify 153, 7, 20, and 4 disks as full, transitional, evolved,
and evolved or transitional, respectively.
Because debris and evolved transitional disks are indistinguishable with 
mid-IR photometry, we have reported those classes as a single category, 
which contains 17 members.

\section{Relative Ages of Populations in Ophiuchus}
\label{sec:ages}

Previous studies have reported evidence for multiple episodes
of star formation within the Ophiuchus complex. 
For instance, \cite{wil05} identified a population of low-extinction stars 
toward L1688 and found that they were older than the embedded population 
and roughly coeval with Upper Sco.
As a result, they proposed that the same event triggered the formation of
the low-extinction stars near L1688 and the members of Upper Sco.
Similarly, \cite{pil16} found a group of stars surrounding the star
$\rho$ Oph that appeared to be older than the embedded stars in L1688
and possibly coeval with Upper Sco.

We can place new constraints on the relative ages of the populations
in Ophiuchus by using {\it Gaia} parallaxes to estimate
absolute magnitudes for the stars in our new census of Ophiuchus.
Evolutionary models predict that low-mass stars evolve 
mostly vertically in the Hertzsprung-Russell during the first $\sim10$~Myr 
after they are born \citep[e.g.][]{bar98}.
As a result, relative values of extinction-corrected absolute magnitudes
($M_K$) at a given spectral type can serve as proxies for relative ages.
We have computed the $M_K$ offsets of members of Ophiuchus from
the median sequence for Upper Sco \citep{luh19} for spectral types of K5--M5,
which spans a range of temperatures in which stars are expected to experience
roughly the same evolution of luminosity with age for 1--10~Myr \citep{bar98}.
Only objects with $\sigma_\pi/\pi \leq 0.1$ from {\it Gaia} DR2 have been
used in these calculations.

\begin{figure}[h]
	\centering
	\includegraphics[trim = 0mm 0mm 0mm 0mm, clip=true, scale=.8]{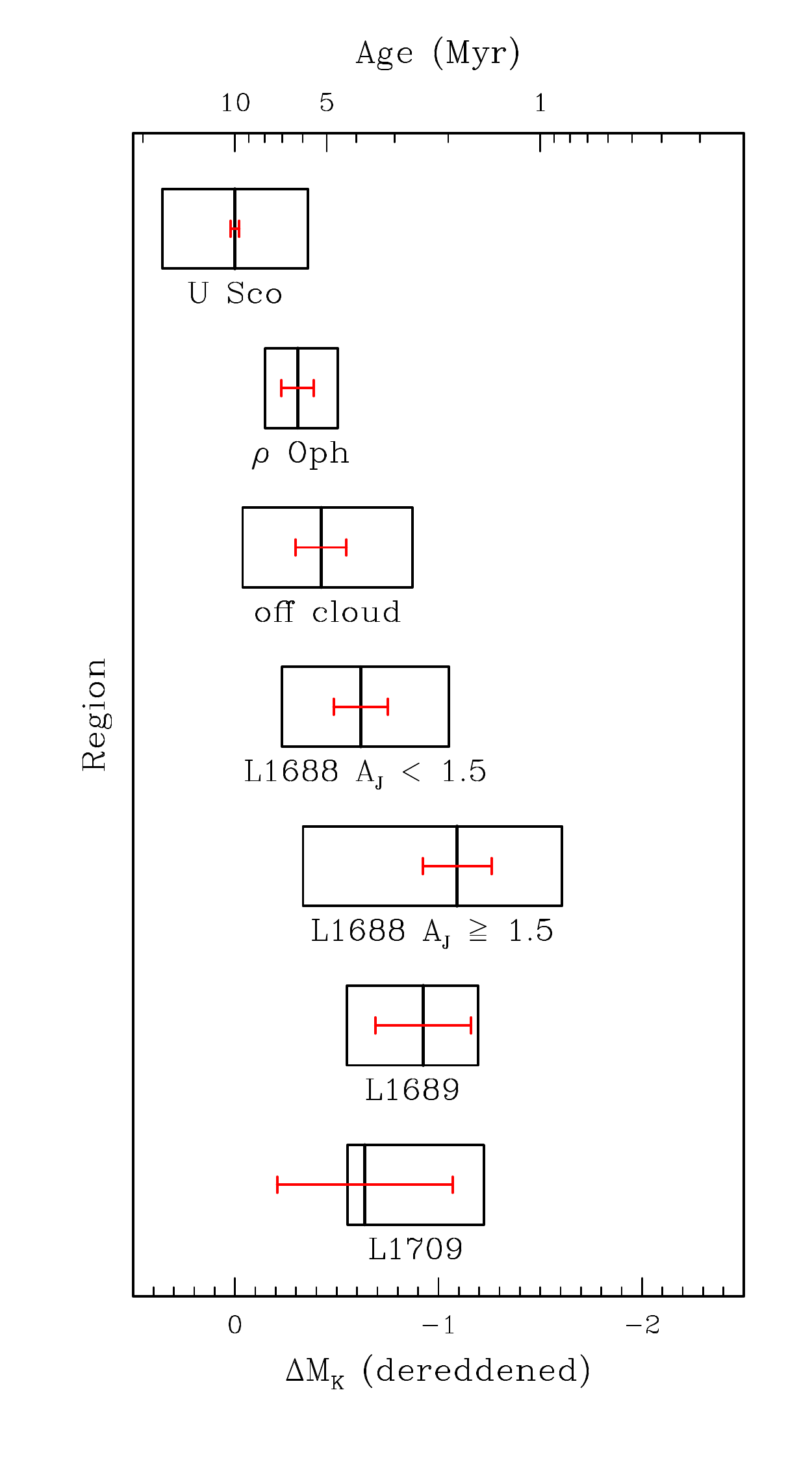}
\caption{
Box diagrams showing the medians (center vertical lines) 
and interquartile ranges (width of box) of the $M_K$ offsets for the
populations in Ophiuchus shown in Figure~\ref{fig:regions}
and for known members of Upper Sco \citep{luh19}. 
Offsets for the K5--M5 members are measured relative to the 
median $M_K$ as a function of spectral type for members of Upper Sco.
Errors for the median values of $\Delta M_K$ have been estimated by
bootstrapping (error bars).
Smaller values of $\Delta M_K$ correspond to brighter photometry, and hence
younger ages.
The top axis indicates the ages that correspond to $\Delta M_K$ assuming
an age of 10 Myr for Upper Sco and the change in luminosity with age
predicted by evolutionary models \citep{bar15}.
}
\label{fig:box}
\end{figure}

In Figure~\ref{fig:box}, we present a box-and-whisker plot of the $M_K$
offsets for the populations of Ophiuchus described in 
Section~\ref{sec:cat} and shown on the map in Figure \ref{fig:regions}.
For comparison, we have included data for known members of Upper Sco. 
Only the inner quartile ranges are plotted. We have
estimated the errors of the median of each population via bootstrapping. 
Because a few of the groups have only a small number of known members, 
those errors can be quite large (e.g., the error of L1709 exceeds the inner 
quartile range).
Along the top axis of Figure \ref{fig:regions}, we have indicated the
ages that correspond to $\Delta M_K$ assuming an age of 10 Myr for Upper Sco
\citep{pec16} and the evolution of luminosity with age predicted 
by evolutionary models \citep{bar15,cho16,dot16}. 
The predicted changes in luminosity with age are similar between the
non-magnetic and magnetic models for the ranges of temperature and ages
in question \citep{fei16}. According to the data in Figure~\ref{fig:box},
the group in L1689 and the embedded stars in L1688 have similar median ages
($\sim2$~Myr) and are the youngest of the samples in Ophiuchus.
L1709 is roughly coeval with those two populations given the errors.
The measurements of $\Delta M_K$ imply progressively older ages for
the low-extinction stars in L1688 (3--4~Myr), the off-cloud stars
($\sim5$~Myr), and $\rho$~Oph ($\sim6$~Myr), none of which are 
as old as Upper Sco.
These results support the conclusion from \citet{pil16} that the $\rho$~Oph
cluster is older than the embedded population in L1688. However, unlike
that study, we find that the former is younger than Upper Sco.
Given that the low-extinction stars toward L1688 and the off-cloud stars
differ only slightly in age, have similar surface densities on the sky,
and have indistinguishable kinematics and extinctions, we propose
that they are members of the same population.
The small age difference between those two samples could be explained
by contamination of the low-extinction stars in front of L1688 by 
stars associated with the younger embedded population that have unusually low
extinctions.

\section{Conclusion}

We have performed a survey for young stars associated with the 
Ophiuchus complex of dark clouds using proper motions and parallaxes from 
{\it Gaia} DR2, proper motions measured with IRAC on \spitz, and 
CMDs constructed with optical and IR photometry from various sources.
We have obtained optical and near-IR spectra of candidates identified with
those data to measure their spectral types and verify their youth,
159 of which are classified as new young objects.  
Based on available measurements of kinematics, most notably from {\it Gaia} DR2,
we adopt 102, 47, and 6 of those sources as members of Ophiuchus,
Upper Sco, and other populations in Sco-Cen, respectively.
Using those same criteria, we also have assessed the membership of all other
known young stars near Ophiuchus, arriving at a catalog of 373 adopted
members of the cloud complex. 
The new Ophiuchus members have spectral types as late as M9.5 
and include 6 of the 10 faintest known members in terms of 
extinction-corrected $K_s$. There remain 259 candidate members of Ophiuchus
from this work that lack sufficient spectroscopic data to assess 
whether they are young.

We have compiled mid-IR photometry from \wise\ and \spitz\ for the adopted
members of Ophiuchus and have used those data to check for excess emission
from circumstellar disks and to classify the evolutionary stages of the
detected disks.
We find that 210 of the members show evidence of disks, including 48 disks
that are in advanced stages of evolution.
In addition, we have estimated the relative median ages of the populations
in Ophiuchus using offsets in $M_K$ from the median sequence of Upper Sco
for members with {\it Gaia} parallaxes and spectral types of K5--M5. 
Those offsets combined with the predicted evolution of luminosity
with age \citet{bar15,cho16,dot16} imply median ages of 
$\sim2$~Myr for L1689 and embedded stars in L1688,
3--4~Myr for low-extinction stars near L1688, and $\sim6$~Myr
for the group containing $\rho$~Oph if we assume an age of 10~Myr
for Upper Sco \citep{pec16}. We suggest that most of the low-extinction stars
toward L1688 may be members of a more widely distributed population that
extends a few degrees to the north and west of that cloud.

\acknowledgements
We thank K. Allers for providing the modified version of Spextool for use with
ARCoIRIS data.
K. L. acknowledges support from NSF grant AST-1208239 and NASA grant
80NSSC18K0444.
The {\it Spitzer Space Telescope} and the IPAC Infrared Science Archive (IRSA)
are operated by JPL and Caltech under contract with NASA.
\wise\ and {\it NEOWISE} are joint projects of
the University of California, Los Angeles, and the Jet Propulsion
Laboratory (JPL)/California Institute of Technology (Caltech), funded by NASA.
2MASS is a joint project of the University of
Massachusetts and the Infrared Processing and Analysis Center (IPAC) at
Caltech, funded by NASA and the NSF.
The IRTF is operated by the University of Hawaii under
contract NNH14CK55B with NASA.
This work has made use of data from the European Space Agency (ESA)
mission {\it Gaia} (\url{https://www.cosmos.esa.int/gaia}), processed by
the {\it Gaia} Data Processing and Analysis Consortium (DPAC,
\url{https://www.cosmos.esa.int/web/gaia/dpac/consortium}). Funding
for the DPAC has been provided by national institutions, in particular
the institutions participating in the {\it Gaia} Multilateral Agreement.
The Center for Exoplanets and Habitable Worlds is supported by the
Pennsylvania State University, the Eberly College of Science, and the
Pennsylvania Space Grant Consortium.
Based in part on observations at Cerro Tololo Inter-American Observatory, National Optical Astronomy Observatory (NOAO Prop. ID: 2016A-0157 \& 2017A-0103
 and PI: T. Esplin), which is operated by the Association of Universities for Research in Astronomy (AURA) under a cooperative agreement with the National Science Foundation. 
Based on observations obtained at the Gemini Observatory (NOAO Prop. ID: 2016A-0139 and PI: T. Esplin), which is operated by the Association of Universities for Research in Astronomy, Inc., under a cooperative agreement with the NSF on behalf of the Gemini partnership: the National Science Foundation (United States), the National Research Council (Canada), CONICYT (Chile), Ministerio de Ciencia, Tecnolog\'{i}a e Innovaci\'{o}n Productiva (Argentina), and Minist\'{e}rio da Ci\^{e}ncia, Tecnologia e Inova\c{c}\~{a}o (Brazil).
Based on observations obtained as part of the VISTA Hemisphere Survey, ESO Progam, 179.A-2010 (PI: McMahon).

{\it Facilities: } 
\facility{Blanco (COSMOS, ARCoIRIS)}, 
\facility{Gemini:South (Flamingos-2)},
\facility{Gemini:North (GNIRS)},
\facility{IRTF (SpeX)},
\facility{{\it Spitzer} (IRAC, MIPS)},
\facility{{\it WISE, Gaia}},
\facility{Magellan:Baade (IMACS)}.

\clearpage

\LongTables
\begin{deluxetable}{ll}
\tabletypesize{\scriptsize}
\tablewidth{300pt}
\tablecaption{Members of the Ophiuchus Complex\label{tab:mem}}
\tablehead{
\colhead{Column Label} &
\colhead{Description}}
\startdata
2MASS    &   2MASS Point Source Catalog source name \\
UGCS      &  UKIDSS Galactic Clusters Survey source name\tablenotemark{a}  \\
WISE      &  WISE source name\tablenotemark{b} \\
Gaia      &  Gaia DR2 source name\\
SR         & Designation from \cite{str49}\\
DoAr      &  Designation from \cite{dol59}\\
GSS        & Designation from \cite{gra73}\\
VSSG       & Designation from \cite{vrb75}\\
Elias     &  Designation from \cite{eli78}\\
ROX    &     Designation from \cite{mon83}\\
WL       &   Designation from \cite{wil83}\\
WSB    &     Designation from \cite{wil87}\\
HBC     &    Designation from \cite{her88}\\
WLY     &    Designation from \cite{wil89}\\
LFAM   &     Designation from \cite{leo91}\\
GY       &   Designation from \cite{gre92}\\
ROXs   &     Designation from \cite{bou92}\\
ROXR1 &      Designation from \cite{cas95}\\
ROXA   &     Designation from \cite{kam97}\\
RXJ      &   Designation from \cite{mar98}\\
ROXR    &    Designation from \cite{gro00}\\
ISO-Oph &     Designation from \cite{bon01}\\
ROXN     &   Designation from \cite{oza05}\\
CFHTWIR-Oph & Designation from \cite{alv10}\\
OName & Other common identifier\\
RAdeg & Right Ascension (J2000) \\
DEdeg & Declination (J2000) \\
Ref-Pos & Reference for RAdeg and DEdeg\tablenotemark{c} \\ 
Region & Region in Figure~\ref{fig:regions} \\
SpType & Spectral Type \\
r\_SpType & Reference for SpType\tablenotemark{d} \\
Adopt & Adopted spectral type \\
GaiapmRA & Proper motion in RA from {\it Gaia} DR2\\
e\_GaiapmRA & Error in GaiapmRA \\
GaiapmDec & Proper motion in DE from {\it Gaia} DR2\\
e\_GaiapmDec & Error in GaiapmDec \\
plx & Parallax from {\it Gaia} DR2\\
e\_plx & Error in plx \\
Gmag & $G$ magnitude from {\it Gaia} DR2\\
e\_Gmag & Error in Gmag \\
GBPmag & $G_{\rm BP}$ magnitude from {\it Gaia} DR2\\
e\_GBPmag & Error in GBPmag \\
GRPmag & $G_{\rm RP}$ magnitude from {\it Gaia} DR2\\
e\_GRPmag & Error in GRPmag \\
RUWE & Re-normalized unit weight error from \citet{lin18} \\
IRpmRA & Proper motion in RA from IRAC \\
e\_IRpmRA & Error in IRpmRA \\
IRpmDec & Proper motion in DE from IRAC \\
e\_IRpmDec & Error in IRpmDec \\
Jmag & $J$ magnitude \\
e\_Jmag & Error in Jmag \\
r\_Jmag & Reference for Jmag\tablenotemark{e}\\
Hmag & $H$ magnitude \\
e\_Hmag & Error in Hmag \\
r\_Hmag & Reference for Hmag\tablenotemark{e}\\
Kmag & $K_s$ or $K$ magnitude \\
e\_Kmag & Error in Kmag \\
r\_Kmag & Reference for Kmag\tablenotemark{e}\\
3.6mag & {\it Spitzer} [3.6] magnitude \\
e\_3.6mag & Error in 3.6mag \\
f\_3.6mag & Flag on 3.6mag\tablenotemark{f} \\
4.5mag & {\it Spitzer} [4.5] magnitude \\
e\_4.5mag & Error in [4.5] magnitude \\
f\_4.5mag & Flag on 4.5mag\tablenotemark{f} \\
5.8mag & {\it Spitzer} [5.8] magnitude \\
e\_5.8mag & Error in 4.5mag \\
f\_5.8mag & Flag on 5.8mag\tablenotemark{f} \\
8.0mag & {\it Spitzer} [8.0] magnitude \\
e\_8.0mag & Error in 8.0mag \\
f\_8.0mag & Flag on 8.0mag\tablenotemark{f} \\
24mag & {\it Spitzer} [24] magnitude \\
e\_24mag & Error in 24mag \\
f\_24mag & Flag on 24mag\tablenotemark{f} \\
W1mag & {\it WISE} $W1$ magnitude \\
e\_W1mag & Error in W1mag \\
f\_W1mag & Flag on W1mag\tablenotemark{f} \\
W2mag & {\it WISE} $W2$ magnitude \\
e\_W2mag & Error in W2mag \\
f\_W2mag & Flag on W2mag\tablenotemark{f} \\
W3mag & {\it WISE} $W3$ magnitude \\
e\_W3mag & Error in W3mag \\
f\_W3mag & Flag on W3mag\tablenotemark{f} \\
W4mag & {\it WISE} $W4$ magnitude \\
e\_W4mag & Error in W4mag \\
f\_W4mag & Flag on W4mag\tablenotemark{f} \\
Exc4.5 & Excess present in [4.5]? \\
Exc8.0 & Excess present in [8.0]? \\
Exc24 & Excess present in [24]? \\
ExcW2 & Excess present in $W2$? \\
ExcW3 & Excess present in $W3$? \\
ExcW4 & Excess present in $W4$? \\
DiskType & Disk Type \\
Ak & Extinction in K \\
f\_Ak & Method for estimating Ak\tablenotemark{g}
\enddata
\tablecomments{This table is available in its entirety in a machine-readable form.}
\tablenotetext{a}{Based on coordinates from Data Release 10 of the UKIDSS Galactic
          Clusters Survey for stars with $K_s > 10$ from 2MASS.}
\tablenotetext{b}{Coordinate-based identifications from the AllWISE Source
          Catalog when available. Otherwise, identifications are from
          the AllWISE Reject Table or the WISE All-Sky Catalog.}
\tablenotetext{c}{
  2MASS = 2MASS Point Source Catalog;
  UKIDSS = UKIDSS DR10 Catalog;
  Gaia = Gaia DR2;
  Kraus14 = \cite{kra14}.
}
\tablenotetext{d}{
1 = \cite{luh19};
2 = \cite{luh18};
3 = \cite{mar98};
4 = \cite{pra07};
5 = This work;
6 = \cite{rom12};
7 = \cite{mcc10};
8 = \cite{esp18};
9 = \cite{gul12};
10 = \cite{all07};
11 = \cite{pre98};
12 = \cite{ria06};
13 = \cite{cie10};
14 = \cite{mer10};
15 = \cite{sle06};
16 = \cite{riz15};
17 = \cite{eri11};
18 = \cite{wil05};
19 = \cite{pra03};
20 = \cite{coh79};
21 = \cite{mah03};
22 = \cite{alv12};
23 = \cite{hou88};
24 = \cite{gre95};
25 = \cite{luh99};
26 = \cite{bou92};
27 = \cite{pec16};
28 = \cite{alv10};
29 = \cite{wal94};
30 = \cite{sta18};
31 = \cite{str49};
32 = \cite{muz12};
33 = \cite{sle08};
34 = \cite{ans16};
35 = \cite{wil99};
36 = \cite{nat02};
37 = \cite{man15};
38 = \cite{cod17};
39 = \cite{tes02};
40 = \cite{com10};
41 = \cite{cus00};
42 = \cite{luh97};
43 = \cite{her14};
44 = \cite{gat06};
45 = \cite{gee07};
46 = \cite{nat06};
47 = \cite{ryd80};
48 = \cite{bow14};
49 = \cite{tor06};
50 = \cite{vrb93};
51 = \cite{whe18}.
}
\tablenotetext{e}{
    2MASS = 2MASS Point Source Catalog;
    UKIDSS = UKIDSS Data Release 10;
    ISPI = our ISPI photometry (Sec~\ref{sec:photo});
    VHS = VHS Data Release 6.
}
\tablenotetext{f}{
nodet = non-detection;
    sat = saturated;
    out = outside of the camera's field of view;
     bl = photometry may be affected by blending with a nearby star;
    bin = includes an unresolved binary companion;
    err = W2 magnitudes brighter than ~6 mag are erroneous;
  unres = too close to a brighter star to be detected;
    ext = photometry is known or suspected to be contaminated by
          extended emission (no data given when extended emission dominates);
  false = detection from WISE catalog appears false or unreliable based
          on visual inspection.}
\tablenotetext{g}{J-H and H-K = derived from these colors assuming
photospheric near-IR colors \citep{luh19};
IR spec = derived from an IR spectrum.}
\end{deluxetable}

\clearpage

\begin{deluxetable}{ll}
\tabletypesize{\scriptsize}
\tablewidth{300pt}
\tablecaption{Young Non-members Toward the Ophiuchus Complex\label{tab:other}}
\tablehead{
\colhead{Column Label} &
\colhead{Description}}
\startdata
2MASS    &   2MASS Point Source Catalog source name \\
UGCS      &  UKIDSS Galactic Clusters Survey source name\tablenotemark{a}  \\
WISE      &  WISE source name\tablenotemark{b} \\
Gaia      &  Gaia DR2 source name\\
SR         & Designation from \cite{str49}\\
DoAr      &  Designation from \cite{dol59}\\
GSS        & Designation from \cite{gra73}\\
VSSG       & Designation from \cite{vrb75}\\
Elias     &  Designation from \cite{eli78}\\
ROX    &     Designation from \cite{mon83}\\
WSB    &     Designation from \cite{wil87}\\
HBC     &    Designation from \cite{her88}\\
WLY     &    Designation from \cite{wil89}\\
LFAM   &     Designation from \cite{leo91}\\
GY       &   Designation from \cite{gre92}\\
ROXs   &     Designation from \cite{bou92}\\
ROXR1 &      Designation from \cite{cas95}\\
ROXA   &     Designation from \cite{kam97}\\
RXJ      &   Designation from \cite{mar98}\\
ROXR    &    Designation from \cite{gro00}\\
ISO-Oph &     Designation from \cite{bon01}\\
OName & Other common identifier\\
RAdeg & Right Ascension (J2000) from Gaia DR2\\
DEdeg & Declination (J2000) from Gaia DR2 \\
SpType & Spectral Type \\
r\_SpType & Reference for SpType\tablenotemark{c} \\
Adopt & Adopted spectral type \\
GaiapmRA & Proper motion in RA from {\it Gaia} DR2\\
e\_GaiapmRA & Error in GaiapmRA \\
GaiapmDec & Proper motion in DE from {\it Gaia} DR2\\
e\_GaiapmDec & Error in GaiapmDec \\
plx & Parallax from {\it Gaia} DR2\\
e\_plx & Error in plx \\
Gmag & $G$ magnitude from {\it Gaia} DR2\\
e\_Gmag & Error in Gmag \\
GBPmag & $G_{\rm BP}$ magnitude from {\it Gaia} DR2\\
e\_GBPmag & Error in GBPmag \\
GRPmag & $G_{\rm RP}$ magnitude from {\it Gaia} DR2\\
e\_GRPmag & Error in GRPmag \\
RUWE & Re-normalized unit weight error from \citet{lin18} \\
Jmag & $J$ magnitude from the 2MASS Point Source Catalog\\
e\_Jmag & Error in Jmag \\
Hmag & $H$ magnitude from the 2MASS Point Source Catalog\\
e\_Hmag & Error in Hmag \\
Kmag & $K_s$ magnitude from the 2MASS Point Source Catalog\\
e\_Kmag & Error in Kmag \\
Pops & Gaia parallax and proper motion consistent with these populations\tablenotemark{d}
\enddata
\tablecomments{This table is available in its entirety in a machine-readable form.}
\tablenotetext{a}{Based on coordinates from Data Release 10 of the UKIDSS Galactic
          Clusters Survey for stars with $K_s > 10$ from 2MASS.}
\tablenotetext{b}{Coordinate-based identifications from the AllWISE Source
          Catalog when available. Otherwise, identifications are from
          the AllWISE Reject Table or the WISE All-Sky Catalog.}
\tablenotetext{c}{
1 = \cite{mar98};
2 = \cite{gul12};
3 = This work;
4 = \cite{hou88};
5 = \cite{esp18};
6 = \cite{wal94};
7 = \cite{gar67};
8 = \cite{her05};
9 = \cite{wil05};
10 = \cite{eri11};
11 = \cite{bou92};
12 = \cite{luh99};
13 = \cite{her14};
14 = \cite{riz15};
15 = \cite{wil99};
16 = \cite{muz12};
17 = \cite{coh79};
18 = \cite{gre95};
19 = \cite{tor06};
20 = \cite{sle08};
21 = \cite{pec16};
22 = \cite{pat93};
23 = \cite{vrb93};
24 = \cite{mcc10};
25 = \cite{cie10}.
}
\tablenotetext{d}{
O/o = Ophiuchus; U/u = Upper Sco; S/s = remainder of Sco-Cen;
n = none of those populations. Upper case letters
indicate that both proper motion and parallax support membership
in that population. Lower case letters indicate that either proper motion
or parallax supports membership while the other parameter is inaccurate,
unreliable, or unavailable.
}
\end{deluxetable}

\clearpage

\begin{deluxetable}{ccll}
\tabletypesize{\scriptsize}
\tablecaption{IRAC Observations of Ophiuchus\label{tab:epochs}}
\tablehead{
\colhead{AOR} & \colhead{PID} & \colhead{PI} & \colhead{epoch}
}
\startdata
 3651840   &  6 &G. Fazio  &     2004.2\\
 3652096   &  6 &G. Fazio   &    2004.2\\
 3660800   &  6 &G. Fazio   &    2004.2\\
 3661056   &  6 &G. Fazio   &    2004.2\\
 5672192   &173 &N. Evans&       2004.6
\enddata
\tablecomments{This table is available in its entirety in machine-readable form. 
 A portion is shown is here for guidance regarding its form and content.}
\end{deluxetable}

\begin{deluxetable}{ll}
\tabletypesize{\scriptsize}
\tablewidth{0pt}
\tablecaption{Candidate Members of Upper Sco and Ophiuchus\label{tab:cand}}
\tablehead{
\colhead{Column Label} &
\colhead{Description}}
\startdata
2MASS & 2MASS Point Source Catalog source name \\
WISEA & AllWISE Source Catalog source name \\
UGCS & UKIDSS Galactic Clusters Survey source name\tablenotemark{a}\\
Gaia & {\it Gaia} DR2 source name \\
RAdeg & Right Ascension (J2000) \\
DEdeg & Declination (J2000) \\
Ref-Pos & Reference for RAdeg and DEdeg\tablenotemark{b} \\
pmRA & Proper motion in right ascension from {\it Gaia} DR2\\
e\_pmRA & Error in pmRA \\
pmDec & Proper motion in declination from {\it Gaia} DR2\\
e\_pmDec & Error in pmDec \\
plx & Parallax from {\it Gaia} DR2\\
e\_plx & Error in plx \\
Gmag & $G$ magnitude from {\it Gaia} DR2\\
e\_Gmag & Error in Gmag \\
GBPmag & $G_{\rm BP}$ magnitude from {\it Gaia} DR2\\
e\_GBPmag & Error in GBPmag \\
GRPmag & $G_{\rm RP}$ magnitude from {\it Gaia} DR2\\
e\_GRPmag & Error in GRPmag \\
RUWE & renormalized unit weight error from \citet{lin18} \\
Jmag & $J$ magnitude \\
e\_Jmag & Error in Jmag \\
r\_Jmag & Reference for Jmag\tablenotemark{c}\\
Hmag & $H$ magnitude \\
e\_Hmag & Error in Hmag \\
r\_Jmag & Reference for Hmag\tablenotemark{c}\\
Kmag & $K_s$ magnitude \\
e\_Kmag & Error in Kmag \\
r\_Kmag & Reference for Kmag\tablenotemark{c}\\
selection & Selection criteria satisfied by candidate\tablenotemark{d} \\
separation & Angular separation from nearest known young star within $5\arcsec$ \\
compGaia & {\it Gaia} DR2 source name of nearest known young star within $5\arcsec$
\enddata
\tablenotetext{a}{Based on coordinates from DR10 of the
UKIDSS Galactic Clusters Survey for stars with $K_s>10$ from 2MASS.}
\tablenotetext{b}{Sources of the right ascension and declination are 
DR2 of {\it Gaia}, DR10 of the UKIDSS Galactic Clusters Survey, 
and the 2MASS Point Source Catalog.}
\tablenotetext{c}{
    2MASS = 2MASS Point Source Catalog;
    UKIDSS = UKIDSS Data Release 10;
    ISPI = our ISPI photometry (Sec~\ref{sec:photo});
    VHS = VHS Data Release 6.
}
\tablenotetext{d}{
G/ip/zp/yp/Z/Y/H/[3.6]/[4.5] = CMDs from Figure~\ref{fig:criteria};
pi = parallax/proper motions from {\it Gaia} DR2;
irac = proper motions from IRAC.}
\tablecomments{
The table is available in its entirety in machine-readable form.}
\end{deluxetable}

\begin{deluxetable}{llll}
\tabletypesize{\scriptsize}
\tablewidth{0pt}
\tablecaption{Observing Log\label{tab:log}}
\tablehead{
\colhead{Telescope/Instrument} &
\colhead{Disperser/Aperture} &
\colhead{Wavelengths/Resolution} &
\colhead{Targets}}
\startdata
IRTF/SpeX & prism/$0\farcs8$ slit & 0.8--2.5~\micron/$R=150$ & 96 \\
Gemini North/GNIRS & 31.7 l~mm$^{-1}$/$1\arcsec$ slit & 0.9--2.5~\micron/$R=600$ & 7 \\
4m Blanco CTIO/COSMOS & red VPH/$0\farcs9$ slit & 0.55--0.95~\micron/3~\AA & 41\\
4m Blanco CTIO/ARCoIRIS & 110.5 l~mm$^{-1}$ + prism/$1\farcs1$ slit &0.8--2.47~\micron/3500 & 56\\
Gemini South/FLAMINGOS-2& HK Grism/$0\farcs72$ slit &1.10--2.65~\micron/450 & 4 \\
Magellan/CorMASS & 40 l~mm$^{-1}$ + prism/$0\farcs4$ slit & 0.8--2.5~\micron/300 & 24\\
Magellan/IMACS & 200 l~mm$^{-1}$ grism/$0\farcs9$ slit & 0.5--1.0~\micron/7~\AA & 4
\enddata
\end{deluxetable}

\begin{deluxetable}{lccccc}
\tabletypesize{\scriptsize}
\tablewidth{0pt}
\tablecaption{Spectroscopic Data for Candidate Members of Ophiuchus and Upper Sco\label{tab:spec}}
\tablehead{
\colhead{Source Name\tablenotemark{a}} &
\colhead{Spectral} & 
\colhead{Instrument} &
\colhead{Date} &
\colhead{Young?} &
\colhead{Adopted Member} \\
\colhead{} &
\colhead{Type} &
\colhead{} &
\colhead{} &
\colhead{} &
\colhead{of Oph?} 
}
\startdata
2MASS J16200698-2430501 &  M5    &        SpeX       &       2016 Apr 8     &          Y & N \\ 
2MASS J16201835-2409537 &  M3.5  &        SpeX       &       2016 Apr 7     &          Y & N \\ 
2MASS J16202629-2416226 &  M3.5  &        ARCoIRIS   &       2016 Jun 19    &          Y & N \\ 
2MASS J16204233-2431473 &  M3    &        ARCoIRIS   &       2016 Jun 18    &          Y & N \\ 
2MASS J16204607-2430599 &  M2.5  &        ARCoIRIS   &       2016 Jun 18    &          Y & N 
\enddata
\tablecomments{This table is available in its entirety in a machine-readable form. A portion is shown is here for guidance regarding its form and content.}
\tablenotetext{a}{Identifications from the 2MASS Point Source Catalog when available.
Otherwise, they based on coordinates from data release 10 of the
UKIDSS Galactic Clusters Survey.}
\end{deluxetable}

 \clearpage

\end{document}